# Gridscape: A Tool for the Creation of Interactive and Dynamic Grid Testbed Web Portals

Hussein Gibbins and Rajkumar Buyya

**Gri**d Computing and **D**istributed **S**ystems (**GRI**DS) Laboratory
Department of Computer Science and Software Engineering
The University of Melbourne, Australia
Email: {hag, raj}@cs.mu.oz.au

**Abstract.** The notion of grid computing has gained an increasing popularity recently as a realistic solution to many of our large-scale data storage and processing needs. It enables the sharing, selection and aggregation of resources geographically distributed across collaborative organisations. Now more and more people are beginning to embrace grid computing and thus are seeing the need to set up their own grids and grid testbeds. With this comes the need to have some means to enable them to view and monitor the status of the resources in these testbeds (eg. Web based Grid portal). Generally developers invest a substantial amount of time and effort developing custom monitoring software. To overcome this limitation, this paper proposes Gridscape – a tool that enables the rapid creation of interactive and dynamic testbed portals (without any programming effort). Gridscape primarily aims to provide a solution for those users who need to be able to create a grid testbed portal but don't necessarily have the time or resources to build a system of their own from scratch.

## 1. Introduction

As we develop into an information dependant society, the amount of information we produce and the amount that we desire to consume continues to grow. As we continue to advance in this information society, we find the need for more sophisticated technology, faster computation and large-scale storage capacity to handle our information wants and needs on demand in real-time. Recently we have seen the rise of utilisation of resources distributed across the Internet (e.g., SETI@Home [1]) in order to solve problems that need large-scale computational resources. This paradigm is popularly known as grid computing [2], which has the potential of being able to deal with large-scale data and compute intensive problems and is opening the door to new innovative computing solutions to problems in both scientific (e.g., bioinformatics[22]) and commercial (e.g., portfolio pricing[23]) fields.

The components that make up computational grids include instruments, displays, computational resources, and information resources that are widely distributed in location and are managed by various organisations. Grid technologies enable large-scale sharing of these resources and in these settings, being able to monitor any resources, services, and computations is challenging due to the heterogeneous nature, large numbers, dynamic behavior, and geographical distribution of the entities in which a user might be interested. Consequently, information services are a vital part of any grid software or infrastructure, providing fundamental mechanisms for monitoring, and hence for planning and adapting application behavior [3].

With the rise in popularity of grid computing, we see an increasing number of people moving towards grid enabling their work and their applications. Many people are now attempting to harness distributed resources and are setting up grid testbeds. Once testbeds have been set up, there is a need for some application or portal to enable the viewing and monitoring of the testbed's status. Gridscape, presented in this paper, aims to assist with this problem by assisting the creation of web based portals as well as making administering these portals an easy process.

The design aims of Gridscape are that it should:

- Allow for the rapid creation of grid testbed portals;
- Allow for simple portal management and administration;
- Provide an interactive and dynamic portal;
- Provide a clear and user-friendly overall view of grid testbed resources; and
- Have a flexible design and implementation such that core components can be leveraged, it provides a high level of portability, and a high level of accessibility (from the browsers perspective).

In the remainder of this paper we further identify the need for Gridscape and how it fits into the grid architecture. We then discuss the design and implementation as well as walking through an example of Gridscape's usage.

## 2. Related Work

As mentioned earlier, there is a definite need for software to monitor testbeds once they are created. However, because of the complex nature of grids, creating useful tools to gather and present resource information is a challenge. There are a number of implementations which currently exist, each with their own strengths and weaknesses. The majority of these efforts are either application specific portals or portal development toolkits. Here we discuss a few representative implementations and compare and contrast their design and development methodologies along with their strengths and weaknesses, clarifying the need for a tool such as Gridscape.

The most basic, and perhaps least useful type of implementation is application specific and uses HTML with static content to present resource status information. This type of implementation does not provide users with up-to-date, real-time feedback about the status of grid resources. This type of monitoring tool is easy to create, however limits the relevance and usefulness of the information provided and is also difficult to maintain or keep updated. Also, these types of portals tend to provide complete, unprocessed grid resource information data, which makes it hard to locate specific characteristic about any given resource, thus severely hindering its usefulness as a monitoring tool. An example of such an implementation is an early version of GRIDView, which is used to monitor the status of the US Atlas Grid Testbed [4].

Another, more sophisticated approach is to use dynamic content within HTML (such as with PHP). This allows for a real-time view of how the grid resources are performing, which is ideal for this type of tool. The NorduGrid [5] Grid Monitor is a good example of this, providing current load information as well as processed and user-friendly MDS information. One feature lacking from this and the previous implementation is the availability of a spatial or geographical view of the resources. It is often useful to be able to have a visual picture of where your resources are located geographically. Again, the downside to this type of implementation is that it is tailored specifically to a particular testbed or particular needs, which means that this monitoring tool cannot be used to monitor any general testbed we may want to monitor.

An even more sophisticated tool can be produced with the use of technology such as Java and Java Applets. This approach has been taken in a number of instances, such as the new GRIDView monitoring tool for the US ATLAS Grid Testbed [6]. However, this implementation doesn't provide the user with immediate and concise information, it is again application specific.

Moving away from the application specific type of portal, we see a number of Grid portal development toolkits. They include GridPort [7], GPDK [9], and Legion Portal [10]. These toolkits assist in the construction of application specific portals; however they operate at a much lower level and aim to provide developers with libraries or low-level interfaces to Grid resources in order to assist in portal creation. For example, GridPort toolkit libraries/interfaces have been utilised in the development of NPACI HotPage[8] portal. Gridscape, on the other hand, requires no explicit programming effort in order to create a testbed portal.

The design approach taken with Gridscape is similar to some degree with grid monitoring software, Map Center [11]. Both implementations provide users with the geographical visualisation of grid resources, and also use HTML with dynamic content. In order to make the portal more interactive both have opted to use JavaScript, as a lightweight alternative to a Java applet.

## 3. Architecture

The architecture of Gridscape and its interaction with various Grid components is shown in Figure 1. Gridscape itself consists of three components: web application, administration tool, and interface to grid information service.

*Web Application*: The web application consists of a customisable template portal which provides an interactive graphical view of resource locations and the ability to monitor its status and details, with the added ability of being able to submit queries to identify resources with specific characteristics.

*Administration Application*: The administration tool provides the user with a simple and user-friendly way of customising and updating their personal testbed portal. Users are able to manage the resources to be used in the portal by adding, removing and editing their details.

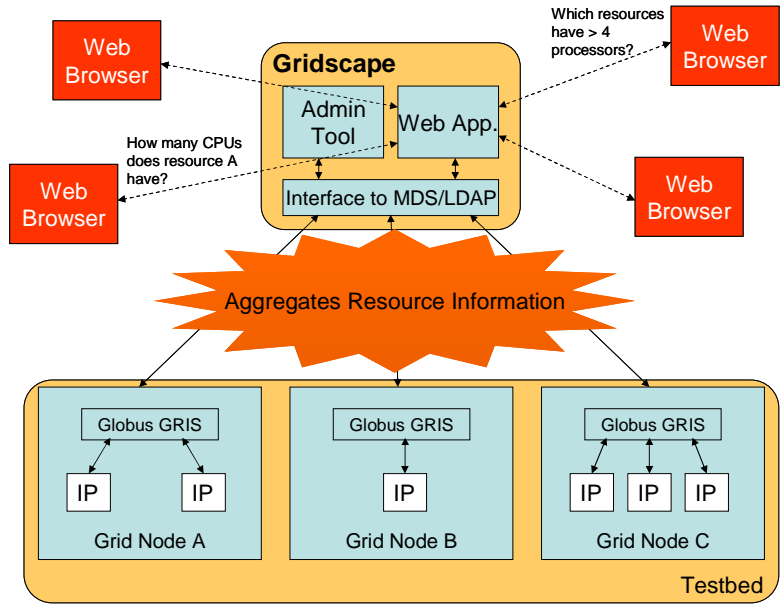

**Figure 1:** Gridscape architecture.

*Interface to Grid Information Service***:** The information provided by Gridscape is gathered from individual grid resources by accessing the Monitoring and Discovery Service (MDS) [3] provided by the Information Services component of the Globus Toolkit [12], which is run on them. MDS is designed to provide a standard mechanism for publishing and discovering resource status and configuration information. It provides a uniform, flexible interface to data collected by lower-level information providers.

Within Gridscape, "interface to MDS" component has been basically developed as a Java based class containing high-level methods that hide low-level details (e.g., LDAP protocols) of accessing MDS services. This level of separation of low-level MDS access mechanisms from other Gridscape components will ensure their portability. For example, if there is a change in MDS access protocols say from LDAP based to XML-based Web services, we can easily update our "MDS access interface" without the need of updating other components.

As the MDS services are utilised by Gridscape while gathering individual Grid node information, it seems logical we first discuss MDS components briefly in order to better understand Gridscape's interaction with them.

**Globus MDS**

The MDS reduces the complexity of accessing system information. This is achieved by having local systems use a wide variety of information-generating and gathering mechanisms, but users only need to know how to interact with MDS to access the information. MDS acts as a point of convergence between the large number of information sources and the large number of applications and high-level services which utilise them.

The MDS represents information in accordance with the Lightweight Directory Access Protocol (LDAP) [3]. LDAP is a set of protocols for accessing information directories. LDAP, a Lightweight version of the old X.500 Directory Access Protocol, supports TCP/IP communication and is becoming the standard protocol when dealing with any directory information applications. Using an LDAP server, MDS provides middleware information in a common interface.

There are three components which make up the MDS hierarchy: Information Providers (IPs), the Grid Resource Information Service (GRIS), and the Grid Index Information Service (GIIS) [13]. At the lowest level there are Information Providers (IPs) which provide resource data such as current load status, CPU configuration, operating system type and version, basic file system information, memory information, and type of network interconnect. These IPs, interface from any data collection service, and report to GRIS. The GRIS runs on a resource and contains the set of information relevant to that resource, provided by the IPs. Individual resources can then be registered to a GIIS, which combines individual GRIS services to provide an overall view of the grid. The GIIS can be explored and searched to find out information about resources, as you would any index.

**Gridscape's Interaction with MDS**

Gridscape discovers the properties of individual resources of a given testbed by making MDS queries to individual GRIS installations. Results are sent back to Gridscape which caches these details for further use. Because Gridscape aims to be free of $3^{rd}$ party tools such as a database, and because querying distributed resources continually is very costly, Gridscape caches the current status of the testbed and allows this store to be shared by any web browsers accessing the portal. The current status information held by Gridscape can be automatically updated periodically, or an immediate status update can be requested at any time.

The action of accessing the GRIS to collect details of individual resources allows Gridscape to behave as a GIIS, to a certain extent, in that it provides users with a collection of separate GRIS information from various resources, in order to provide more of a holistic view of a grid testbed.

## 4. Design and Implementation

The Gridscape web application is designed following the MVC (Model-View-Controller) based, Model-2 type architecture [14] shown in Figure 2. This architecture, which was developed for use with web applications implemented in technology such as Java Server Pages and Servlets, provides a means of decoupling the logic and data-structures of the application (server-side business logic code) from the presentation components (web pages)[15][16]. In order to make implementation easier, and enhance reliability, the Jakarta STRUTS framework [17] has been adopted. STRUTS provides a framework for building Model-2 type web applications.

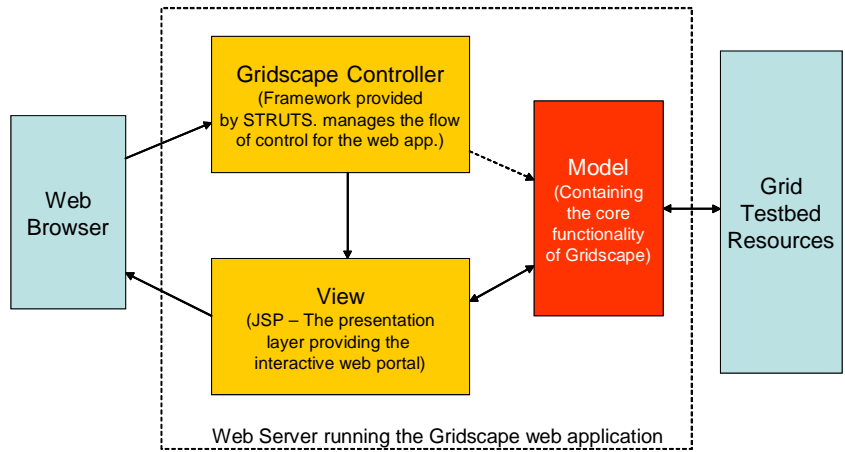

**Figure 2:** MVC Model2 architecture of Gridscape implementation.

With Gridscape, the Model component of the Model-2 architecture becomes the most interesting to investigate further. The reason for this is because this is where the main functionality is located within Gridscape. Also, because of the separation of the presentation, control and the business logic achieved with the application's architecture, we are able to leverage the Model component from the web application and re-use it in the Gridscape Administration Tool. We can see that in this way, other applications could be developed which also make use of the core functionality provided in the Model, by offering a new presentation and control or application layers which access these core components, as illustrated in Figure 3 .

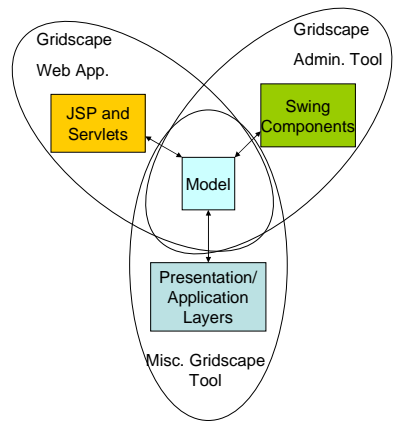

**Figure 3:** Flexibility and reuse of the Model component.

## 4.1 The Model

So far we have identified the significance of the Model and seen its flexibility. The Model itself though consists of a number of important classes. In this section we will take a closer look at some of these classes, their properties, and how they interact with one another.

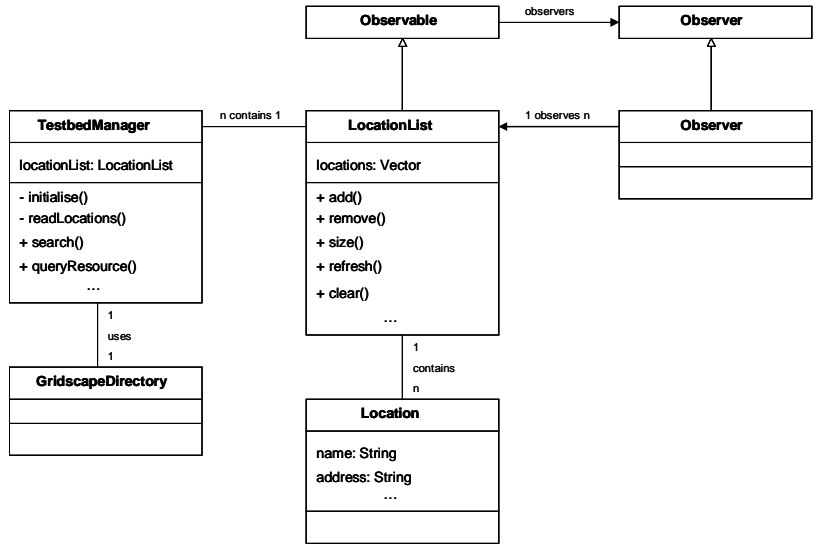

**Figure 4:** Class diagram of the core classes of Gridscape's Model.

### 4.1.1 *GridscapeDirectory*

This class provides a convenient wrapper around the necessary elements of the naming and directory access packages of the core Java API. It provides us with an easy means of connecting to, and querying, the resources in the testbed. It is easy to see that this class will be invoked whenever we communicate with the testbed resources, whether it be through the web application or the admin tool.

### 4.1.2 *Location*

This low level class is used to represent and hold information about a particular resource. This information includes things such as the name, address, port number, as well as the MDS data which is gathered from the resource.

### 4.1.3 *LocationList*

As its name suggests, this class is used to hold a list of the various locations in a testbed. This class extends the Observable class provided by Java. Allowing this class to become observable means, that developing presentation layers or views which depend on this data, is made easy.

### 4.1.4  *TestbedManager*

This class manages the other components within the Model and is responsible for handling queries which are communicated from components outside of the Model. It handles the initialisation of the core of the application and handles duties such as searching by allowing other components to collaborate. It is interesting to note that while the TestbedManager contains an instance of the LocationList, it contains only a singleton instance. The benefit of this type of implementation is that through the web application, even though each client accessing the application is given a new instance of the TestbedManager, there is only one instance of the data. This means that information retrieved from testbed resources is cached, making the site more responsive, and ensuring that everyone is seeing the same up-to-date data.

## 5. Gridscape in Practice

Gridscape has already been used by a number of virtual organisations to create their Grid testbed portals for visualising and monitoring resources [21]. They include Australian Virtual Observatory and UK AstroGrid Collaboration, Belle Analysis Data Grid (BADG), and our own World-Wide Grid (WWG) testbed. In this section we will walk through the steps involved in creation of portal for your own Grid testbed using Gridscape and illustrate them with an example of creating a portal for WWG.

### 5.1  *Deploying the Gridscape web application:*

To begin using Gridscape the user must first deploy the web application within their Jakarta Tomcat installation and also install the administering tool.

### 5.2  *Creating your portal:*

Creating your own customised testbed portal with Gridscape simply involves customising the blank template portal which is provided with Gridscape. Gridscape supports intuitive GUI (see Figure 5) using which you can supply various elements of the testbed: a testbed logo, a map for displaying physical location of resources, and details of resources that are part of the testbed.

### 5.3  *Customising your portal:*

Most of the details about your testbed are stored in configuration files which can also be edited manually. To make customisation easier, an administrating tool has been provided. Open the template from within the administrating tool to continue with customising your portal. Figure 5 shows a snapshot of Gridscape taken while creating a portal for the WWG testbed.

### 5.3.1  *Changing testbed name, logo and other details:*

These items are all customisable from the 'Options' menu. If a new logo is required for the web portal, for example, we can select the 'Change Logo' option. We are then presented with a dialog box which allows us to browse for a suitable image for our logo. Once the logo is selected we can save the selection and the changes will propagate immediately and directly to the web portal. This means that when we next visit the page, we will be able to see this change. This functionality is the same for all options. A small pop-up window shown in Figure 5 illustrates how one can supply testbed name, logo image file, and portal deployment location to Gridscape.

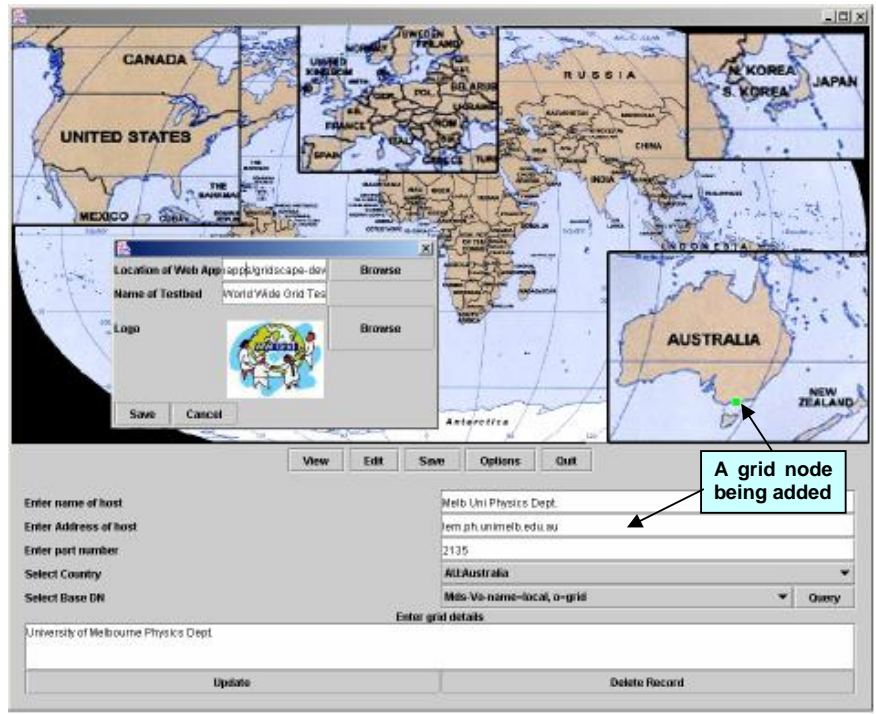

**Figure 5:** A snapshot of Gridscape utilisation while creating the WWG portal.

### 5.3.2   *Managing testbed resources:*

The next step in setting up the portal is to tell Gridscape details about the resources which will be involved. The Gridscape Admin tool provides the user with two modes, an editing and a viewing mode. The editing mode allows the user to edit information regarding resources, while the viewing mode allows the user to simply browse and query existing resources. To add resources to the testbed portal we must enter 'Edit' mode. This can be achieved by choosing the appropriate mode from the 'Mode' menu.

#### 5.3.2.1   *Adding a new resource:*

To add a new resource to the testbed, simply click the mouse in a vacant area on the map. Doing this will automatically create a new resource in your testbed. Position this resource in a desired location on the map by clicking and dragging this resource with the mouse. If the testbed is international, then you need supply a name of the country where the resource is physically located. Figure 5 shows an addition of a Grid node located in the School of Physics at Melbourne University.

#### 5.3.2.2 *Editing resource details:*

When a new resource is created, it is provided with the default property values. To change these properties we first need to select the resource by clicking on it with the mouse. Once selected, we can freely edit such details as its name, address and port number. Once completed, use the 'Update' button to store the changes.

#### 5.3.2.3 *Deleting an unwanted resource:*

If for some reason you need to remove a resource from the testbed, simply select the resource and use the 'Delete' button.

### 5.3.3 *Querying testbed resources:*

Before saving your changes or viewing the web portal online, it is a good idea to go into 'View' mode and query the resources, by clicking on them with the mouse. This will give you confidence that the details you entered were correct and indicate the expected behaviour of your web portal.

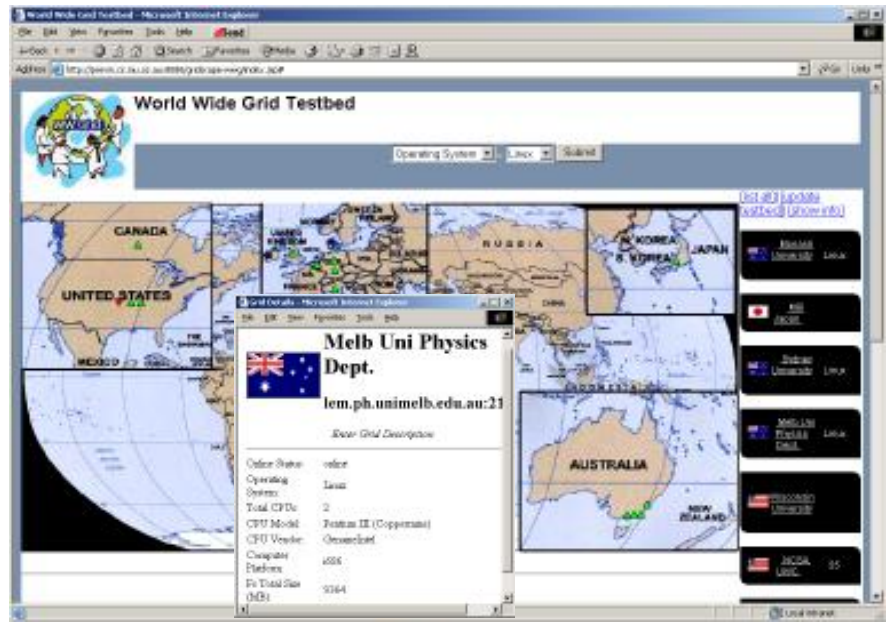

**Figure 6:** A snapshot of browsing World Wide Grid testbed portal.

### 5.4 *Browsing the testbed portal:*

Once the customisation is complete, the testbed details are saved into a configuration file and deployed on the Web server. A snapshot of browsing and monitoring status of the WWG testbed resources through a portal created using Gridscape is shown in

Figure 6. The portal can be accessed online by visiting the World Wide Grid Testbed website [19].

## 6. Conclusion and Future Work

Currently there are a number of unique applications designed to monitor, very specifically, details of only one grid testbed. This paper identifies the need for a tool of this nature - one able to automate the process of creating grid testbed portals. We propose Gridscape, a tool aimed to meet the needs of those who require a testbed portal but simply don't have the resources available to invest in creating their own software from scratch. Gridscape has the potential to provide users with any of the information made available through Globus MDS, and allows for quick and easy creation and administration of web based grid testbed portals.

We are planning to extend Gridscape to support live monitoring of application-level utilisation of Grid resources by integrating it with our Grid application management portal called G-monitor [20].

### *Availability*

The Gridscape software, source, and related information are available for download from the Gridbus project website: http://www.gridbus.org/

An index of sites currently using the Gridscape software can be accessed at the following address: http://previn.cs.mu.oz.au:8080/gridscape/ and the testbed portals can be browsed online.

### *Acknowledgements*

We would like to acknowledge and thank Ankur Chakore, Yogesh Chadee and Rami Safiya, for their contributions in developing the initial monitoring tool called STAMPEDE [18] that served as an early seed for Gridscape. We would like to thank Anthony Sulistio for his comments on early drafts of the paper.